\documentstyle[prl,aps]{revtex}
\input psfig
\def\fr#1/#2{{\textstyle{#1\over#2}}} 

\def\>{\rangle}
\def\<{\langle}
\def\k#1{|\,#1\>}

\def\+{\oplus}
\def\ra{\rightarrow}
\font\ss = cmssbx10
\def\Hc{\hbox{\ss H}}

\def\Xc{\hbox{\ss X}}
\def\Zc{\hbox{\ss Z}}
\def\cX{\hbox{\ss cX}}
\def\cZ{\hbox{\ss cZ}}

\def\c{\fr1/{\sqrt2}}

\def\adj{^\dagger}
\begin{document}

\draft
\twocolumn[\hsize\textwidth\columnwidth\hsize\csname @twocolumnfalse\endcsname
\title{Deconstructing~Dense~Coding}
\author{N.\ David Mermin}
\address{Laboratory of Atomic and Solid State Physics, 
Cornell University,  Ithaca, NY 14853-2501}
\maketitle

\begin{abstract} The remarkable transmission of two bits of
information via a single qubit entangled with another at the
destination, is presented as an expansion of the unremarkable
classical circuit that transmits the bits with two direct qubit-qubit
couplings between source and destination.\end{abstract}  ]


Quantum dense coding\cite{ft:densecoding} enables Alice to communicate
two bits of classical information by sending Bob a single physical
qubit, which is maximally entangled with another qubit already in his
possession. She does the trick by applying one of four unitary
transformations to her member of the entangled pair, thereby
converting the state of the pair into one of four mutually orthogonal
two-qubit states.  Bob can learn which state it is after receiving the
second member of the pair.

What is surprising is that Alice appears to act on only a single
qubit, thereby providing Bob with {\it two\/} bits of information by
sending him only {\it one\/} appropriately prepared qubit.  But this
way of telling the tale downplays a second interaction that takes
place before the curtain even rises on the official story.  That
earlier interaction is required to create the entanglement between the
qubits that Alice and Bob initially share.

The full story remains surprising even with this added prologue, but
the real surprise is that the entangling interaction, essential for
the transmission of the two bits, can take place {\it before\/}
Alice has even chosen the bits she wishes to communicate to Bob.  What
the story really demonstrates is the remarkable ability of entangled
states to store interaction in a highly fungible form that need not be
cashed in until the need arises.

I have made a similar point\cite{ft:telepndm} about quantum
teleportation, showing explicitly how the missing interaction that
makes the difference between a routine classical circuit and a quantum
``miracle'', is buried in the interaction that produces a crucial
shared entangled pair before the state to be teleported need even have
been formed.  Because teleportation and dense coding both exploit
preexisting shared entanglement to facilitate communication with
surprisingly little additional interaction, one might expect there to
be a similar circuit-theoretic deconstruction of dense-coding.  But
because there is no direct mapping from one protocol to the other ---
teleportation involves three qubits and dense coding four --- it is
not obvious from the expansion in \cite{ft:telepndm} of classical
state swapping into quantum teleportation, how dense coding might
arise from an expansion of the classical\cite{ft:classical} circuit
that communicates two bits of information by means of two direct
qubit-to-qubit interactions.  

In this note I show how to do this.  The construction is given in Fig.~1.
The generalization from qubits to $d$-state systems is given in Fig.~2
and Eqs.~(1)-(3).

\begin{figure} \centerline{\psfig{file=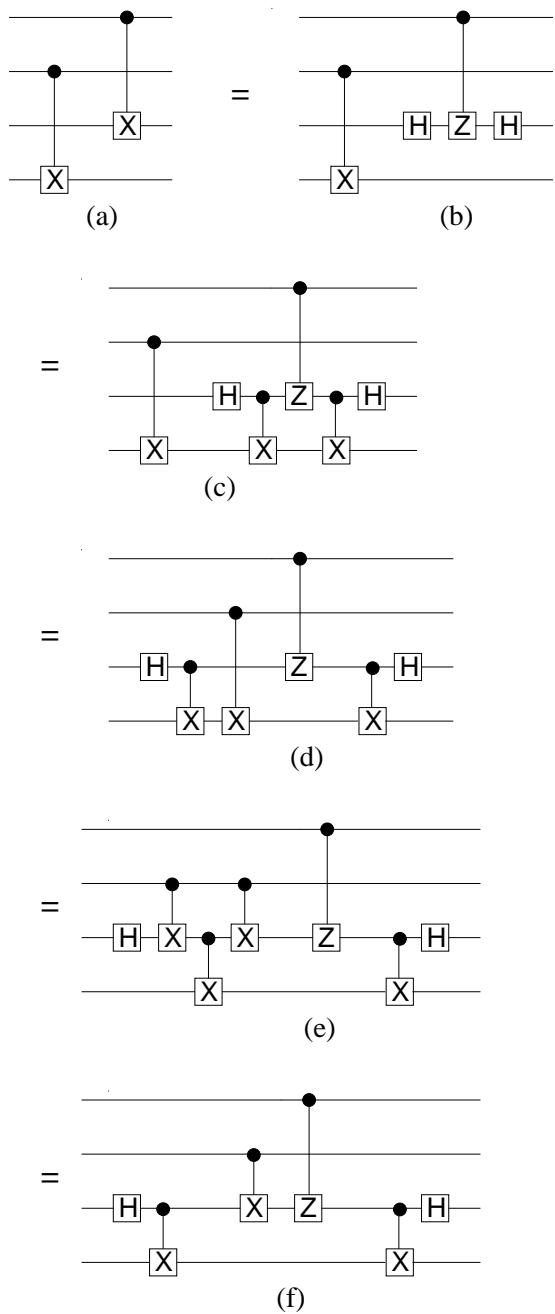,width=3truein}}
\vskip 8pt \caption{How to transform the classical circuit (a) that takes
$\k{xy00}$ to $\k{xyxy}$ by direct couplings within two pairs of
qubits, into the quantum dense-coding circuit (f) that begins with
preparation of an entangled state and ends with a transformation of the
Bell basis into the computational basis.}\end{figure}

If the initial state of the four qubits in Fig.~1(a) is $\k x\k
y\k0\k0$ (reading from top to bottom on the left) then the effect of
the two $\cX$ (cNOT) gates is to transform it into $\k x\k y\k x\k y$.
This automates a classical procedure by which Alice, who possesses the
upper two qubits, can communicate two classical bits of information to
Bob, who possesses the lower two.

To go from this undramatic classically transparent procedure to
quantum dense coding we first expand the $\cX$ on the right into
quantum components, beginning with the fact (Fig.~1(b))
that\cite{ft:qcomp} $\Xc = \Hc\Zc\Hc.$ This is useful because we wish
to eliminate, or at least disguise, the direct coupling on the left
between Alice and Bob's lower qubit.  Because the operator $\Zc$ is
diagonal in the computational basis, it is immaterial whether $\Zc$
acts on a control qubit immediately before or immediately after a
$\cX$.  So since $\cX$ is its own inverse we can expand Fig.~1(b) to
Fig.~1(c), and then move the paired $\cX$ and Hadamard gates to the
extreme left and right, as shown in Fig.~1(d).  The goal of
eliminating the direct coupling between Alice and Bob's lower qubit
can now be achieved by noting that the two $\cX$ gates on the left of
Fig.~1(d) are equivalent to the three $\cX$ gates on the left of
Fig.~1(e), since both sets, acting on the computational basis, leave
the control qubits unaltered, while applying
$\Xc$ to the lowest qubit if and only if the states of the two control
qubits differ.  But since Bob's qubits both start on
the left in the state $\k 0$, and $\Xc$ acts as the identity on
$\Hc\k0 = \c\bigl(\k0+\k1\bigr)$, the leftmost $\cX$ in
Fig.~1(e) always acts as the identity and can be dropped from the
circuit.
 
The result, Fig.~1(f), is an automated dense coding circuit.  The two
gates on the left convert $\k0\k0$ into the maximally entangled state
$\c\bigl(\k0\k0+\k1\k1\bigr).$ The upper member of the entangled pair
is then acted on by $\Xc, \Zc, \Zc\Xc$ or no transformation at all,
depending on whether the state of the upper two qubits is $\k0\k1$,
$\k1\k0$, $\k1\k1$, or $\k0\k0$.  The two gates on the extreme right
then transform the resulting entangled state of the two lower qubits
(one of the four states of the ``Bell basis'') back to whichever
computational basis state of the upper two qubits gave rise to it.

\bigskip

A generalization of the dense-coding protocol from qubits to $d$-state
systems has recently been given by Liu et al\cite{ft:liu}. In the
corresponding generalization of the circuit-theoretic derivation the
$\cX$ operator becomes the controlled bit rotation,
\begin{equation}\cX:\ \ \k x\k y \ra \k x \k{y\+x},\ \ 0 \leq x, y <
d,\label{eq:cX}\end{equation} \noindent where $\+$ denotes addition
modulo $d$, the Hadamard transformation $\Hc$ becomes the quantum
Fourier transform \begin{equation}\Hc: \k y \ra {1 \over \sqrt d}
\sum_{0\leq z<d} e^{2\pi i
zy/d}\k z, \label{eq:Fc}\end{equation}
\begin{figure}
\centerline{\psfig{file=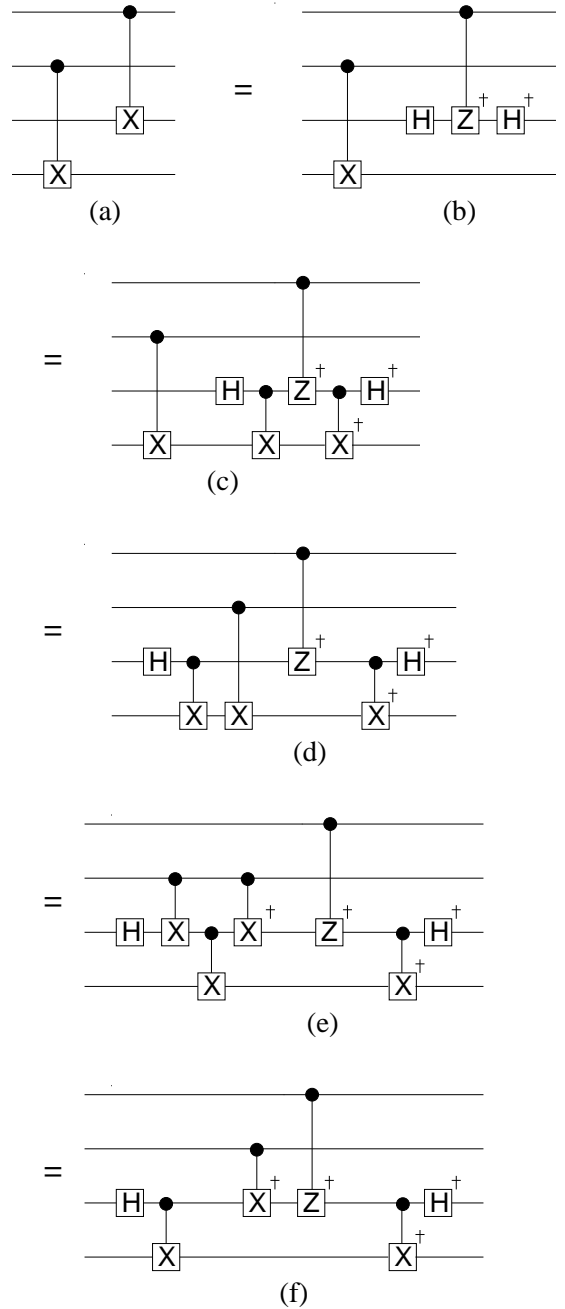,width=3truein}} \vskip 8pt
\caption{The generalizations of $\cX, \cZ,$ and $\Hc$ to $d$-state
systems are no longer their own inverses, but otherwise the extraction
of $d$-state dense coding from the trivial classical circuit is
exactly as in Fig.~1.}\end{figure} 
\hrule
\vskip 10pt
\noindent 
and the controlled-$Z$
operation becomes
\begin{equation}\cZ: \k x \k y \ra e^{-2\pi ixy/d}\k
x\k y. \label{eq:cZ}\end{equation}

\noindent One easily verifies that \begin{equation} (\Hc)_2(\cX)_{12} =
(\cZ)_{12}\adj(\Hc)_2 \label{eq:XZ}
\end{equation} 
and
therefore \begin{equation}\cX_{12} =
(\Hc)_2\adj(\cZ)_{12}\adj(\Hc)_2.\label{eq:FZF}
\end{equation}

Fig.~2 extends the identities of Fig.~1 to $d$-state systems.  The
only difference in the diagrams is that the unitary gates are no
longer self-inverse, and must be distinguished from their
adjoints. Fig.~2(a) shows two direct couplings by controlled bit
rotations (\ref{eq:cX}) that take $\k x\k y\k 0 \k 0$ into $\k x\k y\k
x \k y$, $0 \leq x,y < d$.  Fig.~2(b) introduces\cite{ft:convention}
the identity (\ref{eq:FZF}).  A controlled bit rotation and its
compensating inverse are introduced in Fig.~2(c).  The replacement of
the two $\cX$ gates on the left of Fig.~2(d) by the two $\cX$ and one
$\cX\adj$ gates on the left of Fig.~2(e) is clearly valid for
controlled bit rotations, and the $\cX$ gate on the left of Fig.~2(e)
can be dropped since $\Hc\k0$ is invariant under arbitrary bit
rotations.

Fig.~2(f) is the $d$-state version of dense coding.  The two gates on
the left produce the entangled state \begin{equation} {1 \over \sqrt
d}\sum_{0\leq z < d}\k z\k z.  \label{eq:ent}\end{equation} The two
gates in the middle transform (\ref{eq:ent}) by the action (or
inaction) of the $\cX\adj$ and $\cZ\adj$ gates on the member of the
entangled pair in Alice's possession.  The two gates on the right act
on the pair after both members are in Bob's possession, transforming
its state into that product of Alice's two computational-basis states
that governed the two controlled operations in the middle.

\medskip

These circuit-theoretic deconstructions of dense coding (and the
corresponding deconstructions of teleportation in \cite{ft:telepndm})
back into elementary classical circuits, illustrate the role of
entanglement as interaction-in-advance-of-need, by explicitly tracing
its origin back to a direct classical interaction.  They have the
pedagogical virtue of requiring no algebraic scratchwork whatever
(except for the confirmation of (\ref{eq:FZF}) for $d$-state systems) to
verify that the quantum circuits act as advertised.

\medskip
This work was supported by the National Science
Foundation, Grant PHY0098429.


\begin{references}
\bibitem{ft:classical} I call a circuit classical if every gate
transforms classical states into other classical states --- i.e. if
every gate acts as a permutation on states of the computational basis.
\bibitem{ft:qcomp}I use the standard quantum computational
nomenclature: $\Xc = \sigma_x$, $\Zc = \sigma_z$, and $\Hc =
\c\bigl(\Xc + \Zc)$ is the Hadamard gate.  The two-qubit controlled
gates $\cX$ and $\cZ$ act with $\Xc$ or $\Zc$ on their target qubits
if their control qubits (black dot) are in the state $\k1$ and as the
identity if their control qubits are in the state $\k0$
\bibitem{ft:convention} Note the deplorable but pervasive
convention that the operator on the left acts first in a circuit diagram
even though the operator on the right acts first in an equation.
\end{references}
\end{document}